\PassOptionsToPackage{utf8}{inputenc}
\documentclass{bioinfo}
\copyrightyear{2015} \pubyear{2015}

\access{Advance Access Publication Date: Day Month Year}
\appnotes{Manuscript Category}
\let\cite\citep
\begin{document}
\firstpage{1}

\subtitle{Systems Biology and Networks}

\title[Mitigating cold start problems]{Mitigating cold start problems in drug-target affinity prediction with interaction knowledge transferring}
\author[Sample \textit{et~al}.]{Tri Minh Nguyen\,$^{\text{\sfb 1,}*}$, Thin Nguyen\,$^{\text{\sfb 1}}$ and Truyen Tran\,$^{\text{\sfb 1}}$}
\address{$^{\text{\sf 1}}$Applied Artificial Intelligence Institute,Deakin University,Victoria,Australia\\}

\corresp{$^\ast$To whom correspondence should be addressed.}

\history{Received on XXXXX; revised on XXXXX; accepted on XXXXX}

\editor{Associate Editor: XXXXXXX}

\abstract{
\textbf{Motivation:} Predicting the drug-target interaction is crucial for drug discovery as well as drug repurposing. Machine learning is commonly used in drug-target affinity (DTA) problem. However, machine learning model faces the cold-start problem where the model performance drops when predicting the interaction of a novel drug or target. Previous works try to solve the cold start problem by learning the drug or target representation using unsupervised learning. While the drug or target representation can be learned in an unsupervised manner, it still lacks the interaction information, which is critical in drug-target interaction.\\
\textbf{Results:} To incorporate the interaction information into the drug and protein interaction, we proposed using transfer learning from chemical-chemical interaction (CCI) and protein-protein interaction (PPI) task to drug-target interaction task. The representation learned by CCI and PPI tasks can be transferred smoothly to the DTA task due to the similar nature of the tasks. The result on the drug-target affinity datasets shows that our proposed method has advantages compared to other pretraining methods in the DTA task.\\
\textbf{Availability:} The source code is available at https://github.com/ngminhtri0394/C2P2\\
\textbf{Contact:} \href{minhtri@deakin.edu.au}{minhtri@deakin.edu.au}\\
}

\maketitle
\section{Introduction \label{sec:intro}}
\begin{figure*}[h]
  \includegraphics[width=\textwidth]{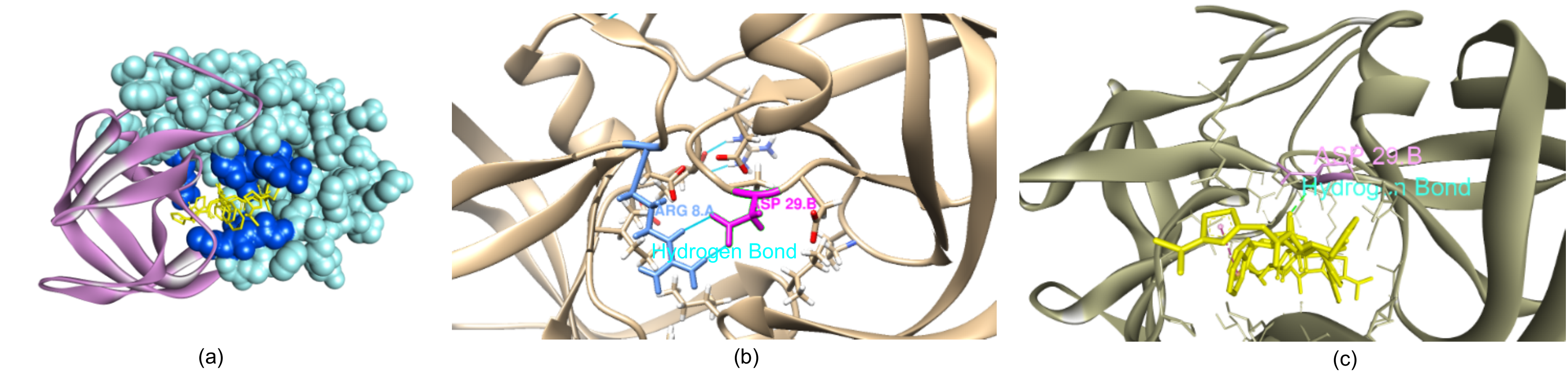}
  \caption{Example of how information from PPI task can be transferred to DTA task. (a) Crystal structure of the complex of resistant strain of HIV-1 protease (v82a mutant) with Ritonavir (b) The hydrogen bond in protein-protein interaction at the protein interface (c) The binding site of Ritonavir in the proximity of protein interface \label{fig:PPIExample}}
\end{figure*}
\begin{figure}
\centering{}\includegraphics[width=0.4\textwidth]{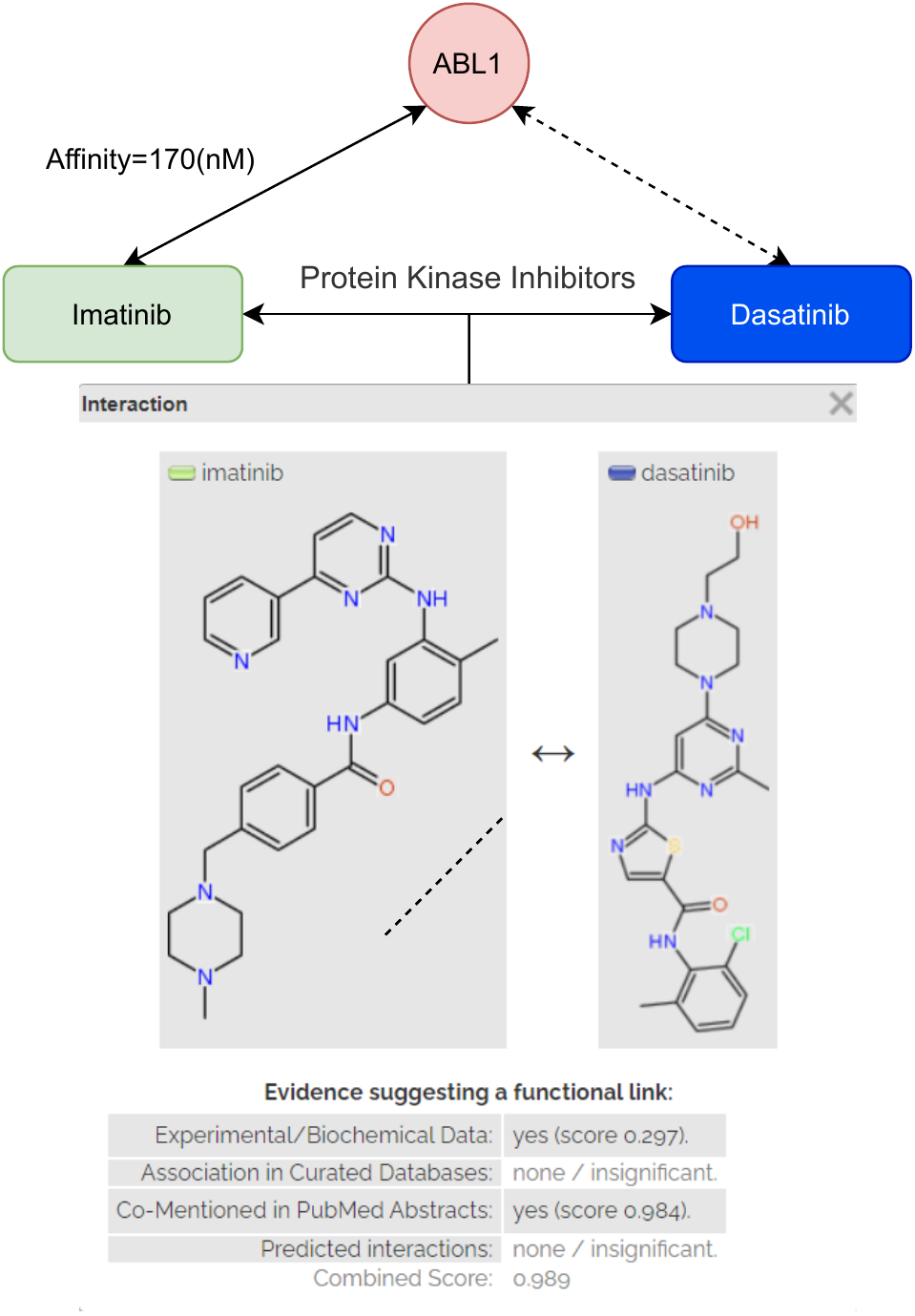} \caption{Chemical-chemical interaction provides external information for drug-target binding. Both Imatinib and Dasatinib share the MeSH pharmacological action 'Protein Kinase Inhibitors' reported in the experimental data of STITCH \cite{Kuhn2008STITCH:Proteins} database. The CCI report is generated by STITCH database web server tool \cite{Kuhn2008STITCH:Proteins}.  
\label{fig:CCI_RLI} }
\end{figure}

Predicting the drug-target interaction is an important task in the drug discovery and drug repurposing \cite{Thafar2019ComparisonAffinities}. Experimental assays provide a precise but expensive tool to determine the binding affinity. On the other hand, computational methods have gained attraction due to their low cost and reasonable performance \cite{Gilson2016BindingDBPharmacology}.

Over the years, many machine learning-based DTA prediction methods \cite{Cichonska2017Computational-experimentalInhibitors,Ozturk2018DeepDTA:Prediction,Nguyen2020GraphDTA:Networks,Nguyen2021GEFA:Prediction} have been proposed. However, these computational methods face the cold-start challenge where the model performance drops in novel drugs or targets, which are common in drug discovery or drug repurposing.

Pre-training is an effective method to handle the cold-start problem. Pre-training helps the model to learn a robust and generalized representation by tapping into a huge amount of unlabeled and labeled data from other relevant tasks. Because both chemicals and proteins can be represented as sequences, language modeling is one of the common pre-training tasks. Thanks to the huge available unlabelled dataset, the model can learn the internal structure arrangement, or in short, the grammar of molecules and proteins by predicting the masked tokens in the sequences. Other pre-training methods such as pre-training graph neural networks, contrastive learning can be either share the same principle as the language model or use different schemes such as mutual information. All the unsupervised pre-training methods share the common strategy which exploits the relationship among components of the structure or between structure classes. These components can vary significantly across atoms, residues, or functional groups. These relationships between components can help the model to learn the meaningful representation of each token as well as the whole sequence. 

Even though the unsupervised pre-training can model the intra-molecule interaction within the molecule or protein to provide the contextual information in the representation, it still lacks the inter-molecule interaction information. By saying inter-molecule interaction, we mean the interaction between the molecule or protein with other entities. Because the essence of the drug-target interaction is in the inter-molecule interaction, it raises the question of whether the intra-molecule interaction information learned by the language model is sufficient for the DTA task. 

To incorporate the inter-molecule interaction into the protein or molecule, we propose a transfer learning framework called \textbf{C}hemical-\textbf{C}hemical \textbf{P}rotein-\textbf{P}rotein Transferred DTA (\textbf{C2P2}). First, C2P2 transfers the inter-molecule interaction knowledge learned from chemical-chemical interaction (CCI) and protein-protein interaction (PPI). Then we combine the inter-molecule interaction with the intra-molecule interaction knowledge to learn the drug-target interaction space. 

Protein-protein interaction is the physical interaction between two or more protein macro-molecules. This interaction is the result of the electrostatics forces, hydrogen bonding, or hydrophobic effect of the residues at the protein interface \cite{Jones1996PrinciplesInteractions}. The properties of the protein interface such as size and shape, complementary between surfaces, residue interface propensities, hydrophobicity, segmentation, secondary structure, and structure flexibility \cite{Jones1996PrinciplesInteractions}. Even though the protein interface is usually viewed as large, flat, featureless, and usually described as undruggable \cite{Hopkins2002TheGenome,Blundell2006StructuralDiscovery,Blundell2000Protein-proteinSignalling}, the protein-protein interaction can reveal the effective drug-target binding mode \cite{Fry2015TargetingDiscovery}. Previous works have taken advantage of PPI in drug discovery \cite{Fry2015TargetingDiscovery,C.Fry2012Small-moleculePartner,Arkin2014Small-MoleculeReality}. In addition, the distribution of the protein interface can indicate the distribution of ligand-binding pocket. Previous work \cite{Gao2012TheFormation} shows that in the protein-protein complex, the majority of ligand binding pocket are with 6 Amstrong (\AA) of the protein interface. Looking at Fig. \ref{fig:PPIExample}, the hydrogen bond between ARG8 and ASP29 in the protein-protein complex (Fig. \ref{fig:PPIExample}b) also exists in the binding configuration with Ritonavir. Therefore, the information from the protein-protein can be beneficial for the drug-target interaction. 

Chemical-chemical interaction (CCI) is the interaction between two chemical entities. The interaction can be derived from various ways such as pathway databases, text mining, structure or activities similarity \cite{Kuhn2008STITCH:Proteins}. CCI can provide information for many related tasks such as toxicity, combination therapies effect, biological functions, and drug-target bindings \cite{Kwon2019End-to-endPrediction}. CCI network can provide information to speed up the drug discovery process \cite{Chen2015TheCancer}. In addition, we can formulate the residue-ligand interaction as a chemical-chemical interaction in which the interaction is the hydrogen bonding, Van der Waals force, or electrostatics (Fig. \ref{fig:CCI_RLI}). In this case, the information from the CCI task can be beneficial for learning the residue-ligand interaction, thus protein-ligand interaction. 

Our contribution is two folds. First, we propose enhancing the drug-target interaction prediction framework with not only inter-molecule interaction learned from language modeling but also intra-molecule interaction learned from related tasks. Second, we provides different ways to integrate the learned intra-molecules information into different representations. 
\section{Related works}
\subsection{Learning protein representation}
\subsubsection{Sequence representation}
Recent developments \cite{Devlin2019BERT:Understanding,Liu2019RoBERTa:Approach} in natural language processing allow the learning model to capture the contextual relationship between tokens in the sequence from a large amount of unlabeled sequence data to achieve state-of-the-art performance on many tasks. The success of the language modeling approach is transferred to protein sequence modeling. TAPE \cite{Rao2019EvaluatingTAPE} learns the protein embedding using language model Transformer \cite{Devlin2019BERT:Understanding} with thirty-one million sequences from the Pfam dataset \cite{El-Gebali2019The2019}. Rives et al.\cite{Rives2021BiologicalSequences} train the language model varying in size in the same manner as TAPE on 250 million sequences of UniRef \cite{Suzek2015UniRefSearches} dataset. ProtTrans \cite{Elnaggar2021ProtTrans:Computing} uses auto-regressive models (Transformer-XL, XLNet) and auto-encoder models (BERT, Albert, Electra, T5) to learn the protein embedding from 2.1 billion protein sequences. 
\subsubsection{3D structure representation}
In the sequential representation, the structure information is lost. Another way to represent the protein is using the exact 3D structure information, meaning using the 3D coordinate to represent each residue. However, acquiring the protein folding information through experimental methods such as X-ray can be time-consuming or expensive. Therefore, several computational methods are proposed \cite{Jumper2021HighlyAlphaFold,Kim2004ProteinServer} to compute high-resolution protein structures. The predicted 3D structure can be used to construct the detailed protein surface using point cloud \cite{Dai2021ProteinLearning} or multi-scale graph structure \cite{Somnath2021Multi-ScaleProteins}. However, predicting the atom's coordinate with high accuracy requires large computational resources. In addition, encoding the whole protein structure to the atom level may lead to sparse representation and inefficient computational resource usage. Therefore, more simple representation can be beneficial.
\subsubsection{Protein graph representation}
To balance between 3D structural information and simplicity, 2D representation via attributed graph can be used. Previous works \cite{Nguyen2021GEFA:Prediction,Jiang2020DrugtargetMaps} have been using protein structure graph representation for DTA prediction. The contact/distance map is used as the adjacency matrix of an attributed graph where each node represents a residue and edge represents the contact/distance between residues. The node attribute can be simply a one-hot encoding of residue type \cite{Jiang2020DrugtargetMaps} or an embedding vector of the residue obtained from the language model \cite{Nguyen2021GEFA:Prediction}.   
\subsection{Learning molecule representation}
\subsubsection{Sequence representation}
The molecules can be represented as SMILES sequence. Therefore, we can apply language modeling to learn the embedding of the molecules. Recent works \cite{Winter2019LearningRepresentations,Chithrananda2020ChemBERTa:Prediction} uses LSTM and Transformer to learn the SMILES sequence representation of chemical space from over 77 million SMILES sequences of PubChem dataset \cite{Kim2019PubChemData}. Chemical SMILES language modeling is essentially an atom level pre-training where the model can learn the intra-interaction of the molecule. 
\subsubsection{Graph representation}
Graph is the natural representation of the molecule in which the atoms are nodes and bonds are edges. Pre-training method on graph neural network allows the model to capture the robust representation at atom level and molecules level. On node level pre-training, Weihua et al. \cite{Hu2020StrategiesNetworks} propose both node-level pre-training via attribute masking and context prediction task and graph-level pre-training via transfer learning from graph attribute and graph structure prediction. On graph level pre-training, InfoGraph \cite{Sun2020InfoGraph:Maximization} maximizes the mutual information between supervised and unsupervised representation. Node level pre-training can help the model to learn the intra-interaction and internal structure at atom level while graph level pre-training allows the model to learn a robust representation of graph structure within the same molecule class.    
\section{Methods}
Drug-target binding affinity (DTA) problem is predicting the binding affinity A between a drug compound D and a protein P. Mathematically, the DTA prediction problem can be formulated as a regression task:
\begin{equation}
A=\mathcal{F}_{\theta}(P,D),
\end{equation}
where $\theta$ is model parameters of predicting function $\mathcal{F}$.

In this section, we present our framework to combine the intra-molecule interaction from language modeling with the inter-molecule interaction knowledge learned from PPI and CCI tasks. In Sec. \ref{sec:overallframework}, we present the overall framework of C2P2, followed by learning inter-molecule and intra-molecule interaction with language modeling, CCI, and PPI task. Then Sec. \ref{sec:dtappicci} introduces the combination of the inter-molecule and intra-molecule interaction to predict the binding affinity.  

\subsection{Overall framework \label{sec:overallframework}}
\begin{figure*}[h]
  \includegraphics[width=\textwidth]{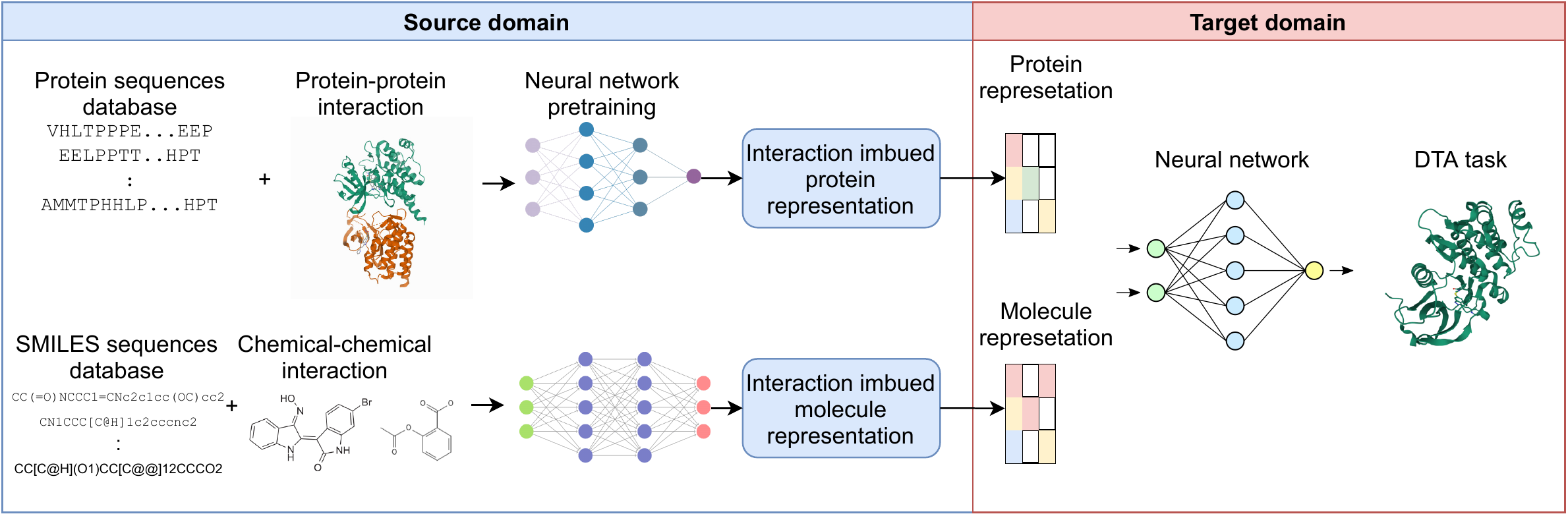}
  \caption{The workflow of transferring the interaction knowledge learned from the source domain of PPI and CCI to the target DTA domain. Pre-training the neural network with CCI and PPI task allows the model to learn the interaction mechanism which may not exist in the current DTA database. \label{fig:highlevel} }
\end{figure*}

\begin{figure}
\centering{}\includegraphics[width=\columnwidth]{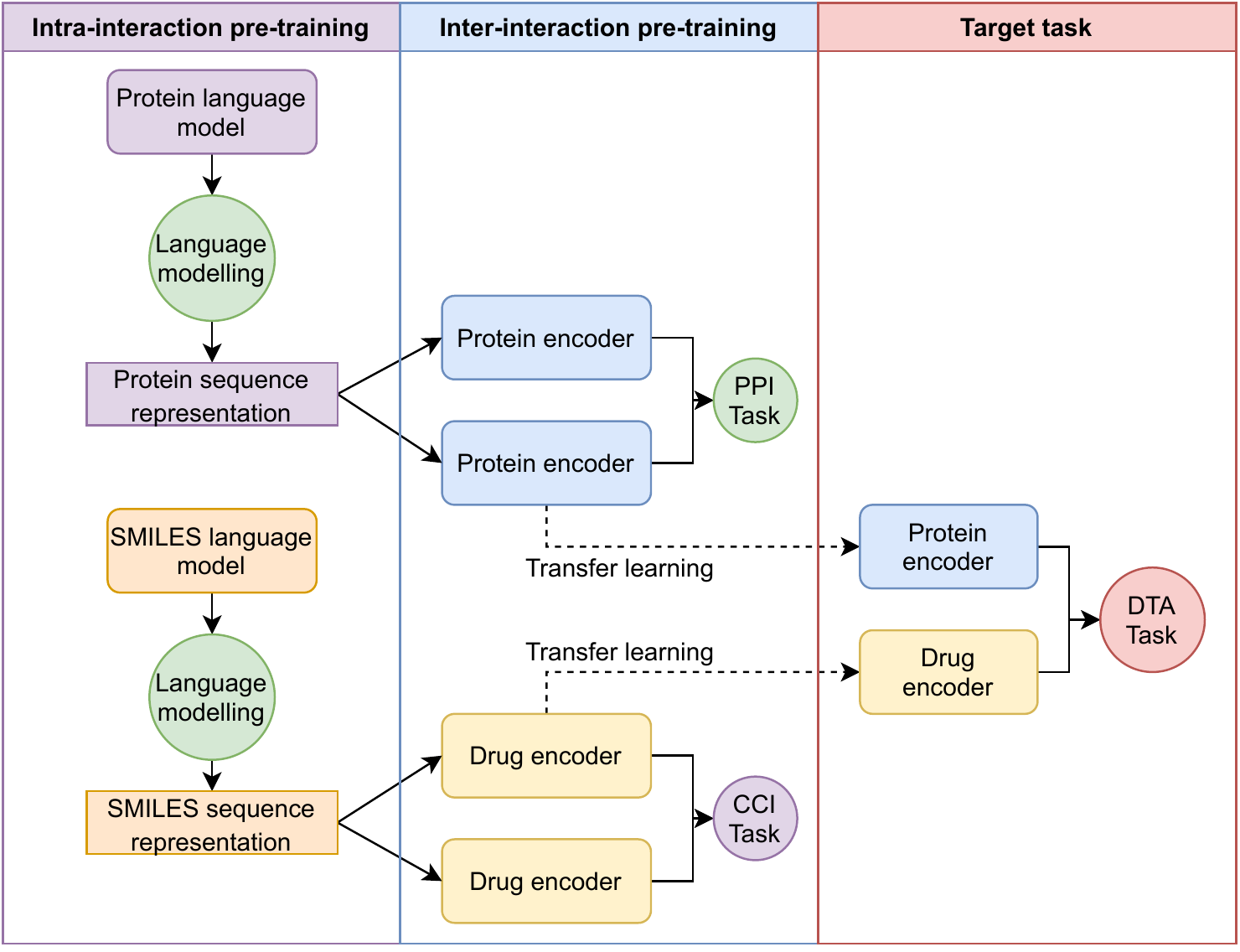} \caption{The framework architecture of the C2P2 model. First, the protein encoder and drug encoder are trained with PPI task and CCI task respectively. Then pre-trained encoders are used for drug and target encoding in the DTA model.  
\label{fig:overall} }
\end{figure}

The overall framework is presented in Fig. \ref{fig:highlevel} and \ref{fig:overall}. The goal is to transfer the interaction learned from the source domain, which is PPI and CCI task, to the target domain DTA task. First, the protein and drug encoder is pre-trained with PPI and CCI tasks. The benefits of pre-training the protein and drug encoder with PPI and CCI tasks are two folds: better generalization representation and interaction-oriented representation. By better generalization representation, we mean that the encoder can learn from a large amount of drug and protein samples from PPI and CCI task. Interaction-oriented representation means that the encoder can learn the binding interaction of many different drugs and proteins. Then the pre-trained drug and target encoders are transferred to the target domain DTA task to extract the drug and target interaction-oriented representation. Finally, both drug and target representation are combined to predict the binding affinity.

\subsection{Learning chemical inter-molecule interaction space \label{sec:cci}} \label{Sec:learningcci}
In this section, we propose the framework to learn the chemical inter-molecule interaction via the chemical-chemical interaction (CCI) prediction task. The overall framework consists of two main steps: learning molecule representation and interaction inference. Our CCI model takes two chemical SMILES sequences $D_{s1}$ and $D_{s2}$ as the inputs. The molecule representations of two SMILES sequences can be either graph representations (Sec. \ref{sec:molgraph}) or language model representations (Sec. \ref{sec:mollm}). Then both representations of $D_{s1}$ and $D_{s2}$ are joined for chemical-chemical interaction. By learning the chemical-chemical interaction, our goal is pre-training the molecule encoder to encode the interaction imbued molecule representation.
\subsubsection{Graph representation of drug molecule \label{sec:molgraph}}
\begin{figure}
\centering{}\includegraphics[width=0.4\textwidth]{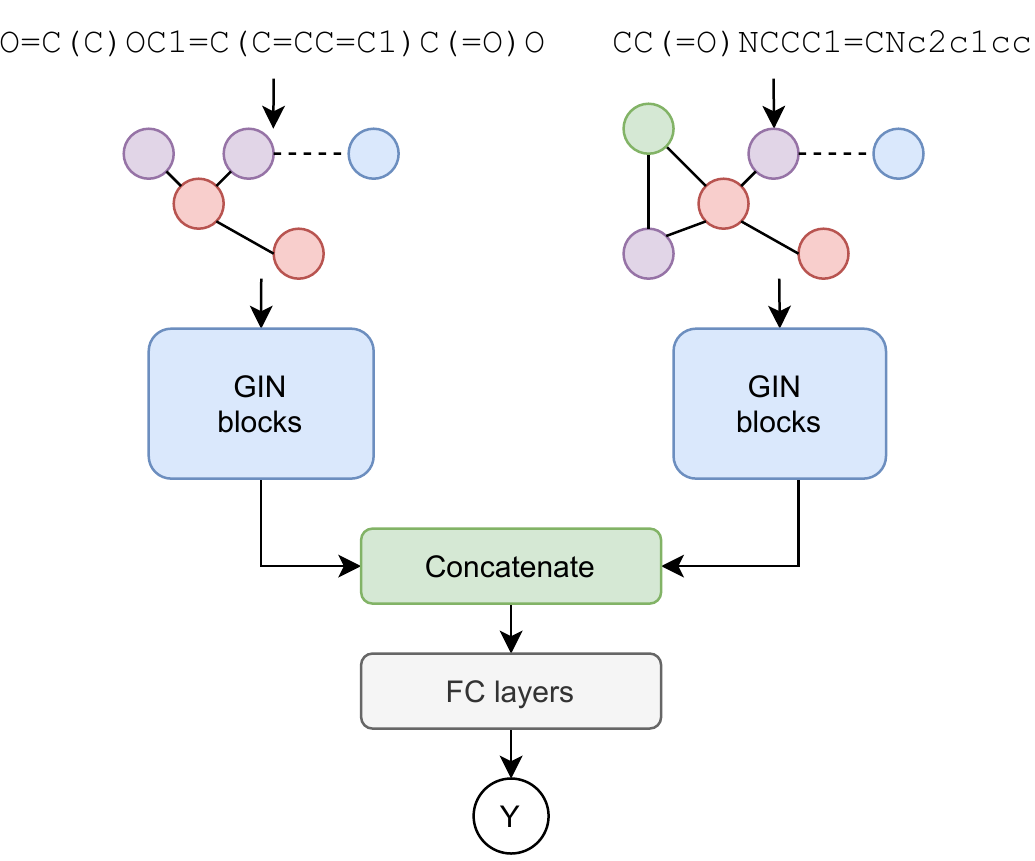} \caption{Learning molecule interaction space with CCI task and Graph encoder
\label{fig:GIN_CCI} }
\end{figure}

Fig. \ref{fig:GIN_CCI} shows the architecture of CCI task with graph neural network. Our CCI framework takes the graph structure $\mathcal{G}_{1}$ and $\mathcal{G}_{2}$ of two molecules. The molecule graph structure has nodes representing the atoms and edges representing the bonds. 
\begin{equation}
    \mathcal{G}=(\mathcal{X},\mathcal{A})
\end{equation}
where $\mathcal{X}={p_1,...,p_v}$ is node feature matrix of $N$ nodes and $\mathcal{A}\in \mathbb{R}^{N\times N}$ is the adjacency matrix that describes the graph structure.

The atom node feature $\mathcal{X}$ is its element type, degree, number of Hydrogens, and implicit valence. The detail of the feature vector of the molecule graph node is shown in Table \ref{tab:graphnodefeat}. The graph representation is learned using Graph Isomorphism Network (GIN) \cite{Xu2018HowNetworks}. The graph neural network updates the node feature vector by:

\begin{equation}
    a_v^{(k)} = AGGREGATE^{(k)}({p_v^{(k-1)} :u \in \mathcal{N}(v)})\\
\end{equation}
\begin{equation}
    p_v^{(k)} = COMBINE^{(k)}(p_v^{(k-1)}, a_v^{(k)})\\
\end{equation}
where $p_v^{(k)} \in \mathbb{R}^{C^{(k)}}$ is the $k$-th layer feature vector of $v$-th node. GIN updates the feature vector $p_v^{(k)}$ by:

\begin{equation}
    p_v^{(k)}=MLP^{(k)}((1+\epsilon^{(k)})\dots p_v^{(k-1)}+\sum_{u\in \mathcal{N}(v)} p_u^{(k-1)}) \in \mathbb{R}^{C^{(k)}}
\end{equation}
where $\epsilon^{(k)}$ is a trainable parameter, MLP is a multi-layer perceptron. 

After $k$-th GIN layers, we have the $\mathcal{P}_{d}^{\prime}=\{p_{i}^{\prime}\mid p_{i}^{\prime}\in \mathbb{R}^{h_{1}}\}_{i=1}^{S}$ as node features of molecule graph, where $S$ is the number of nodes in the drug graph, $h_{1}$ is the dimension of node feature vector. Then we use the max pooling operation followed by linear layer for feature projection:
\begin{equation}
p_{\textrm{max}}^{\prime}=\textrm{MaxPool}(\mathcal{P}_{d}^{\prime}),
\end{equation}
\begin{equation}
x_{d}=(W_{0}p_{\textrm{max}}^{\prime}+b_{0})W_{1}+b_{1}.
\end{equation}
Finally, we obtain $x_{d}$ as the feature vector of the drug molecule. 

\subsubsection{Molecule SMILES representation by language modeling \label{sec:mollm}}

\begin{figure}
\centering{}\includegraphics[width=0.3\textwidth]{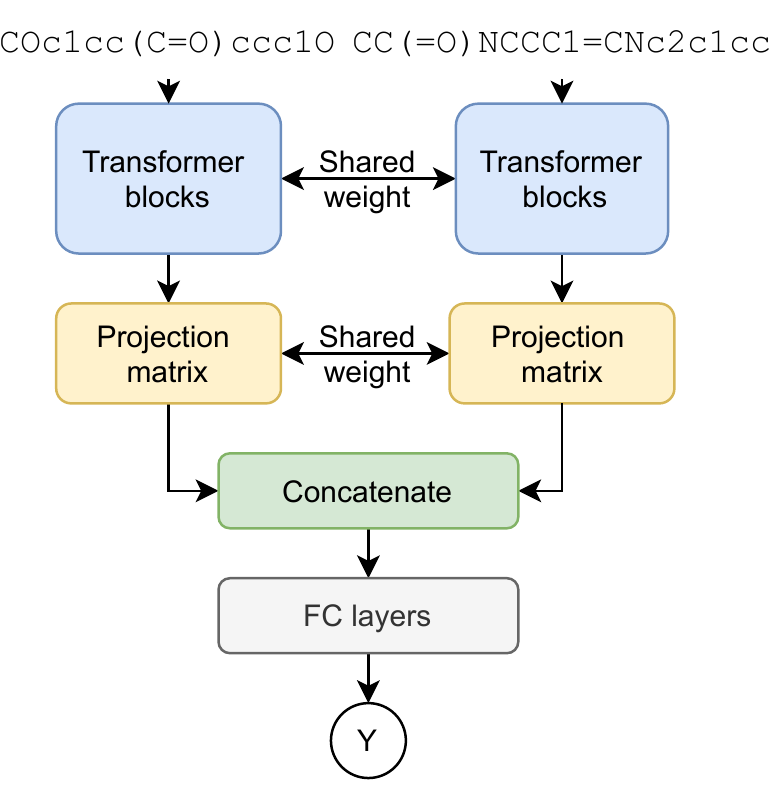} \caption{Enhancing the drug representation learned from language model with interaction information from CCI task
\label{fig:Chemberta_CCI} }
\end{figure}

Fig. \ref{fig:Chemberta_CCI} shows the architecture of enhancing the molecule representation learned from the language model with the interaction information. As the language model tends to learn the internal arrangement (grammar structure) which is essentially the internal interaction. To enhance the language model representation with molecule inter-molecule interaction information, we fine-tune the language model on the CCI task. 

Given the SMILES sequence $D_s$ with length $n$, SMILES sequence representation is extracted using the pre-trained Transformer blocks. We use the BERT language model named Chemberta pre-trained on SMILES sequence \cite{Chithrananda2020ChemBERTa:Prediction}.
\begin{equation}
    X_s = BERT(D_s), X_s \in \mathbb{R}^{n \times d}
\end{equation}
where $d$ is the dimension of the embedding vector.  The pre-training task is predicting the masked character in SMILES sequence. Then the sequence feature vector $x_s$ is the average along feature vector:
\begin{equation}
    x_s = AVG(X_s), x_s \in \mathbb{R}^d
\end{equation}
Then the sequence representation $x_s$ is projected into lower dimension using linear layer:
\begin{equation}\label{eq:projectchemberta}
    x_d = (W_{\theta_{d}}x_s+b), x_d \in \mathbb{R}^{d^{\prime}}, d^{\prime} < d
\end{equation}
The goal of the linear layer is to learn to extract important features from the sequence representation and reduce noise. The Transformer and projection matrix in both branches are shared weight to reduce the number of parameters. 

\subsubsection{Chemical-chemical prediction}
The SMILES sequences from two chemical $D_s1$ and $D_s2$ are encoded into $x_{d1}$ and $x_{d2
}$ by either graph neural network (Sec. \ref{sec:molgraph}) or pre-trained language model (Sec. \ref{sec:mollm}). Then both chemical representations are joined with a simple concatenate operator:
\begin{equation}
    x_{dj} = [x_{d1};x_{d2}]
\end{equation}
Finally, the interaction is predicted with a classifier:
\begin{equation}
    y = sigmoid(RELU(Wx_{df}+b))
\end{equation}

\begin{table}
\caption{Molecule feature vector \label{tab:graphnodefeat}}
\begin{center}
\begin{tabular}{c c } 
\hline
Feature & Feature length \\ 
\hline
Element types& 43\\
Degree&10\\ 
Number of Hydrogens&10\\
Implicit valence&10\\
Aromatic&1\\
\hline
\end{tabular}
\end{center}
\end{table}

\subsection{Learning protein inter-molecule interaction space \label{sec:ppi}}
\subsubsection{Protein sequence representation by language modeling}
\begin{figure}
\centering{}\includegraphics[width=0.3\textwidth]{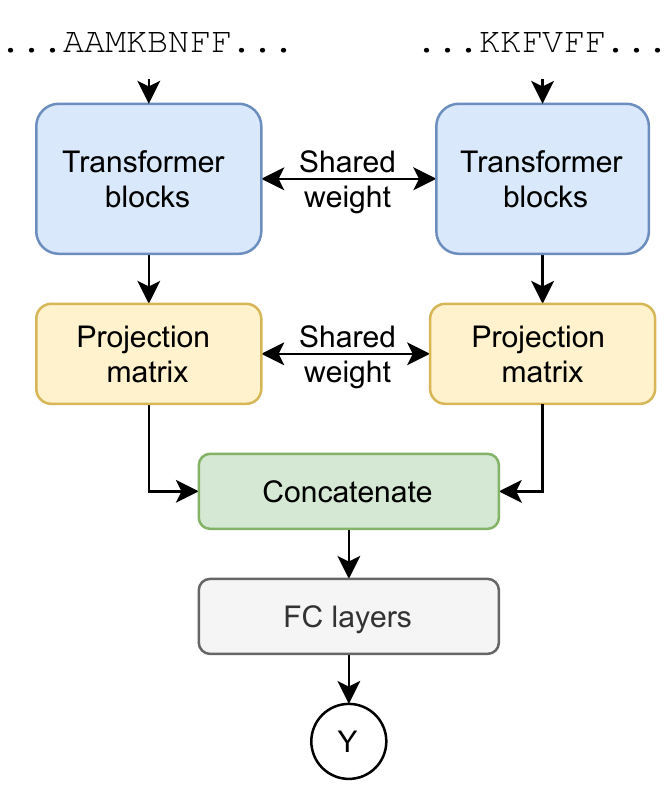} \caption{Enhancing the protein representation learned from language model with interaction information from PPI task.
\label{fig:ESM_PPI} }
\end{figure}

Fig. \ref{fig:ESM_PPI} presents the protein-protein interaction prediction model. The goal is to enhance the protein sequence representation learned by the language model with the protein interaction. Given two protein sequences $D_{p1}$ and $D_{p2}$ length $n$, the protein sequence embedding $X_p$ is extracted by a protein language model named ESM \cite{Rives2021BiologicalSequences}. 

\begin{equation}
    X_p = ESM(D_p), X_p \in \mathbb{R}^{n \times d}
\end{equation}
where $d$ is the embedding dimension. ESM is pre-trained with predicting masked tokens in the protein sequence. The protein sequence embedding is averaged along dimension d:
\begin{equation} \label{eq:projectesm}
    x_p = AVG(X_p), x_p \in \mathbb{R}^d, x_p \in \mathbb{R}^{d^{\prime}}, d^{\prime} < d
\end{equation}

The protein sequence representation $x_p$ is projected into lower dimension using linear layer:
\begin{equation}
    x_p = (W_{\theta_{p}}x_s+b).
\end{equation}
\subsubsection{Protein-protein interaction prediction}
Given the two protein sequence representations $x_{p1}$ and $x_{p2}$ of two input protein sequences $x_{p1}$ and $x_{p2}$, the joint representation is:
\begin{equation}
    x_{pj} = [x_{p1};x_{p2}]
\end{equation}
where $[;]$ is the concatenate operator. The $p1-p2$ interaction is predicted by:
\begin{equation}
    y = sigmoid(RELU(Wx_{pf}+b))
\end{equation}

\subsection{Integrating inter-molecule interaction into DTA model \label{sec:dtappicci}}
After being pre-trained with CCI (Sec. \ref{sec:cci}) and PPI task (Sec. \ref{sec:ppi}), the drug encoder $f(D_s,\theta_d)$ and protein encoder $f(D_p,\theta_p)$, where $\theta_p$ and $\theta_d$ are model parameters, are used to encode the protein and drug for DTA task:
\begin{equation}
    x_p = f(D_p,\theta_p)
\end{equation}
\begin{equation}
    x_d = f(D_s,\theta_d)
\end{equation}
The protein-drug joint representation is:
\begin{equation}
    x_{pdj} = [x_{p};x_{d}]
\end{equation}
Finally, the binding affinity is predicted by:
\begin{equation}
    y_{a}=(W_{0}x_{pdj}+b_{0})W_{1}+b_{1}
\end{equation}

\section{Experiments}
\subsection{Dataset}
We use the STRING dataset \cite{Szklarczyk2021TheSets} for the PPI task. The STRING dataset is the protein-protein network database from over 67.6 million proteins with over 20 billion protein-protein pairs. The protein-protein association includes text mining from literature, interaction experiments, computational experiments, and systematic interaction transferring. As we only need the protein physical interaction, we filter out other types of protein-protein association such as text mining. 

For the CCI task, we use the STITCH dataset \cite{Kuhn2008STITCH:Proteins}. The dataset contains over 0.5 million chemicals with over 1.6 billion interaction. The chemical-chemical associations are built from the experimental results from pathway dataset, text mining from literature, structural similarity, and activities similarity. The drug encoder is pre-trained by either full STITCH dataset or only experimental association. 

For the DTA task, we conduct our experiments on two popular DTA datasets: Davis \cite{Davis2011ComprehensiveSelectivity} and Kiba \cite{Tang2014MakingAnalysis}. In the DTA task, we test our proposed method in cold-start settings, including cold-drug and cold-target.  
\subsection{Benchmark}
We use four benchmark methods to evaluate the performance of extra-interaction transfer learning on different representations. First, we compare our proposed method with the previous SOTA method GraphDTA \cite{Nguyen2020GraphDTA:Networks}. GraphDTA uses CNN as protein encoder and graph neural network as drug encoder. Then the second benchmark method is ESMDTA which replaces the CNN protein encoder with protein representation pre-trained with protein language model ESM \cite{Rives2021BiologicalSequences}. The third benchmark is ChembertaDTA which replaces the graph encoder with SMILES sequences language model representation \cite{Chithrananda2020ChemBERTa:Prediction}. Finally, to evaluate with other graph pre-training strategies, we compare our method with Infograph pre-training method \cite{Sun2020InfoGraph:Maximization}. We evaluate the model performance on the test set using Root Mean Squared Error (RMSE), Pearson \cite{Benesty2009NoiseProcessing}, Spearman \cite{Zwillinger1999CRCFormulae}, and Concordance Index (CI) \cite{Gonen2005ConcordanceRegression}.






\subsection{Implementation detail}
\begin{table}[h!]
\caption{Hyper-parameters in the experiments.\label{tab:hyperparam}}
\begin{center}
\begin{tabular}{c c} 
\hline
Hyper-parameters & Value \\ 
\hline
Learning rate&[0.0005:0.005]\\
Batch size&[128; 256; 512; 1024]\\
\hline
\end{tabular}
\end{center}
\end{table}

Our methods are implemented using Pytorch. The hyper-parameters are tuned using the validation set. The hyper-parameters detail reported in Table.\ref{tab:hyperparam}. The results are reported on the independent test set. The protein language model ESM embedding dimension is $d=768$ which is later projected to $d'=128$ (Eq. \ref{eq:projectesm}). The Chemberta embedding dimension $d=768$ is projected to $d'=128$ (Eq. \ref{eq:projectchemberta}). The model is trained with MSE loss using Adam optimizer for 500 epochs. The number of GIN layers (Sec. \ref{sec:molgraph}) $k=5$. 
\section{Results and Discussion}
\subsection{\textbf{Inter-molecule interaction knowledge benefits the DTA task}}

\begin{table}[h!]
\caption{The performance of the different drug and protein encoder combinations on Davis dataset with the \textbf{Cold-target} setting. The X-Y drug or protein encoder means that base model is X and pre-trained with Y task. PPI, CCI, and Infograph are pre-training with PPI, CCI task, or Infograph unsupervised training. \label{tab:coldtargetdavis}}
\begin{center}
\resizebox{\columnwidth}{!}{%
\begin{tabular}{p{1.5cm} p{1.5cm} c c c c} 
\hline
Drug encoder & Protein encoder & RMSE & Pearson & Spearman & CI \\ 
\hline
GIN(\cite{Nguyen2020GraphDTA:Networks})&CNN(\cite{Nguyen2020GraphDTA:Networks})&0.7102&0.545&0.4507&0.7397\\
GIN&ESM&0.6974&0.5725&0.4838&0.7586\\
&ESM-PPI&\textbf{0.6753}&\textbf{0.5881}&\textbf{0.5241}&\textbf{0.7805}\\
GIN-CCI&ESM&0.6794&0.5891&0.4867&0.7607\\
&ESM-PPI&\textbf{0.6793}&\textbf{0.595}&\textbf{0.4928}&\textbf{0.763}\\
Chemberta&ESM&0.6992&0.5812&0.4816&0.7575\\
&ESM-PPI&\textbf{0.6672}&\textbf{0.6089}&\textbf{0.5361}&\textbf{0.7873}\\
Chemberta-CCI&ESM&\textbf{0.6651}&\textbf{0.6009}&\textbf{0.515}&\textbf{0.7768}\\
&ESM-PPI&0.6697&0.5966&0.4923&0.7632\\
GIN-Infograph&ESM&\textbf{0.6689}&\textbf{0.5923}&\textbf{0.5271}&\textbf{0.7821}\\
&ESM-PPI&0.6852&0.5728&0.4621&0.7459\\
\hline
\end{tabular}
}
\end{center}
\end{table}

\begin{table}[h!]
\caption{The performance of the different drug and protein encoder combinations on PDBBind dataset with the \textbf{Cold-target} setting. The X-Y drug or protein encoder means that base model is X and pre-trained with Y task. PPI, CCI, and Infograph are pre-training with PPI, CCI task, or Infograph unsupervised training.\label{tab:coldtargetpdbbind}}
\begin{center}
\resizebox{\columnwidth}{!}{%
\begin{tabular}{p{1.5cm} p{1.5cm} c c c c} 
\hline
Drug encoder & Protein encoder & RMSE & Pearson & Spearman & CI \\ 
\hline
GIN(\cite{Nguyen2020GraphDTA:Networks})&CNN(\cite{Nguyen2020GraphDTA:Networks})&1.597&0.595&0.5949&0.7116\\
GIN&ESM&1.5385&0.6667&0.6753&0.7472\\
&ESM-PPI&\textbf{1.4320}&\textbf{0.6970}&\textbf{0.6823}&\textbf{0.7499}\\
GIN-CCI&ESM&1.4429&0.6946&0.681&0.7507\\
&ESM-PPI&\textbf{1.3815}&\textbf{0.7152}&\textbf{0.7071}&\textbf{0.7606}\\
Chemberta&ESM&1.3937&0.7177&0.7095&0.7599\\
&ESM-PPI&\textbf{1.3739}&\textbf{0.7241}&\textbf{0.7154}&\textbf{0.7634}\\
Chemberta-CCI&ESM&\textbf{1.3698}&\textbf{0.7175}&\textbf{0.7051}&\textbf{0.7598}\\
&ESM-PPI&1.3806&0.7137&0.7022&0.7579\\
GIN-Infograph&ESM&1.5029&0.6884&0.6862&0.7528\\
&ESM-PPI&\textbf{1.4027}&\textbf{0.7143}&\textbf{0.7094}&\textbf{0.7616}\\
\hline
\end{tabular}
}
\end{center}
\end{table}

\begin{table}[h!]
\caption{The performance of the different drug and protein encoder combinations on Davis dataset with the \textbf{Cold-drug} setting. The X-Y drug or protein encoder means that base model is X and pre-trained with Y task. PPI, CCI, and Infograph are pre-training with PPI, CCI task, or Infograph unsupervised training. \label{tab:colddrugdavis}}
\begin{center}
\resizebox{\columnwidth}{!}{%
\begin{tabular}{p{1.5cm} p{1.5cm} c c c c} 
\hline
Protein encoder & Drug encoder & RMSE & Pearson & Spearman & CI \\ 
\hline
CNN(\cite{Nguyen2020GraphDTA:Networks})&GIN(\cite{Nguyen2020GraphDTA:Networks})&0.9485&0.413&0.3998&0.6903\\
ESM&GIN-Infograph&0.9614&0.4919&0.4964&0.7413\\
&GIN&0.9853&0.3579&0.4416&0.7088\\
&GIN-CCI&\textbf{0.8755}&\textbf{0.575}&\textbf{0.5034}&\textbf{0.743}\\
&Chemberta&0.9169&0.5174&0.3974&0.6909\\
&Chemberta-CCI&\textbf{0.9146}&{0.5259}&\textbf{0.4485}&\textbf{0.7171}\\
ESM-PPI&GIN-Infograph&0.9637&0.4535&0.4206&0.7007\\
&GIN&0.9489&0.4202&0.4027&0.692\\
&GIN-CCI&\textbf{0.8841}&\textbf{0.5564}&\textbf{0.4741}&\textbf{0.7299}\\
&Chemberta&\textbf{0.9032}&\textbf{0.4935}&0.3449&0.6645\\
&Chemberta-CCI&0.9171&0.4906&\textbf{0.4216}&\textbf{0.7034}\\
\hline
\end{tabular}
}
\end{center}
\end{table}

\begin{table}[h!]
\caption{The performance of the different drug and protein encoder combinations on PDBBind dataset with the \textbf{Cold-drug} setting. The X-Y drug or protein encoder means that base model is X and pre-trained with Y task. PPI, CCI, and Infograph are pre-training with PPI, CCI task, or Infograph unsupervised training.\label{tab:colddrugpdbbind}}
\begin{center}
\resizebox{\columnwidth}{!}{%
\begin{tabular}{p{1.5cm} p{1.5cm} c c c c} 
\hline
Protein encoder & Drug encoder & RMSE & Pearson & Spearman & CI \\ 
\hline
CNN(\cite{Nguyen2020GraphDTA:Networks})&GIN(\cite{Nguyen2020GraphDTA:Networks})&1.5263&0.6237&0.6116&0.7193\\
ESM&GIN-Infograph&1.3618&0.7152&0.6943&0.7564\\
&GIN&1.3600&0.7172&0.693&0.755\\
&GIN-CCI&\textbf{1.3484}&\textbf{0.7236}&\textbf{0.7025}&\textbf{0.7603}\\
&Chemberta&1.3962&0.6878&0.6624&0.7405\\
&Chemberta-CCI&1.3653&0.7059&0.6798&0.7498\\
ESM-PPI&GIN-Infograph&1.3857&0.703&0.6871&0.7534\\
&GIN&1.359&0.7124&0.6922&0.7559\\
&GIN-CCI&\textbf{1.3379}&\textbf{0.7282}&\textbf{0.7039}&\textbf{0.7618}\\
&Chemberta&1.3652&0.7183&0.6943&0.756\\
&Chemberta-CCI&1.3735&0.7009&0.68&0.75\\
\hline
\end{tabular}
}
\end{center}
\end{table}

We demonstrate the advantages of transferring the inter-molecule interaction learned from PPI and CCI tasks to the DTA tasks in cold-drug and cold-target settings across two benchmark datasets with balance distribution (PDBBind dataset) and long-tail distribution (Davis dataset). 

In the cold-target setting, we group the proposed methods by the drug encoder and compare the performance between models with and without PPI transfer learning. Overall, models with PPI transfer learning show advantages compared with model without transfer learning. With the graph-based drug encoder (GIN, GIN-CCI, and Infograph), PPI enhanced models has better overall performance compared to model using only ESM feature. Looking at the language model-based drug encoder, the combination of Chemberta as drug encoder and ESM-PPI as protein encoder consistently outperforms model with only ESM as protein encoder. However, combining Chemberta-CCI with ESM feature outperforms ESM-PPI feature across two datasets. This suggests some degree of incompatibility between Chemberta-CCI and ESM-PPI in the cold-target setting. In the end, in general, cooperating the intra-molecule information learned from PPI task with protein language model such as ESM benefits the DTA task performance.

Similar to the cold-target setting, for the cold-drug setting, we group the proposed models by protein encoder and compare the performance of models with and without CCI transfer learning. Among graph-based drug encoders, pretraining graph neural network with CCI task outperforms Infograph pretraining and training from scratch across two datasets and two types of protein encoder. In case of language model-based drug encoder, while pairing with ESM protein encoder, models with CCI pretraining have better performance than models without pre-training. However, Chemberta-CCI and ESM-PPI show a certain degree of incompatibility shown in lower performance than Chemberta and ESM-PPI pair. Overall, integrating CCI information into DTA models enhances the DTA model performance, especially in graph representation.   

\subsection{\textbf{Protein-protein interaction knowledge enhances protein language model representation}}

\begin{figure*}
\centering{}\includegraphics[width=0.7\textwidth]{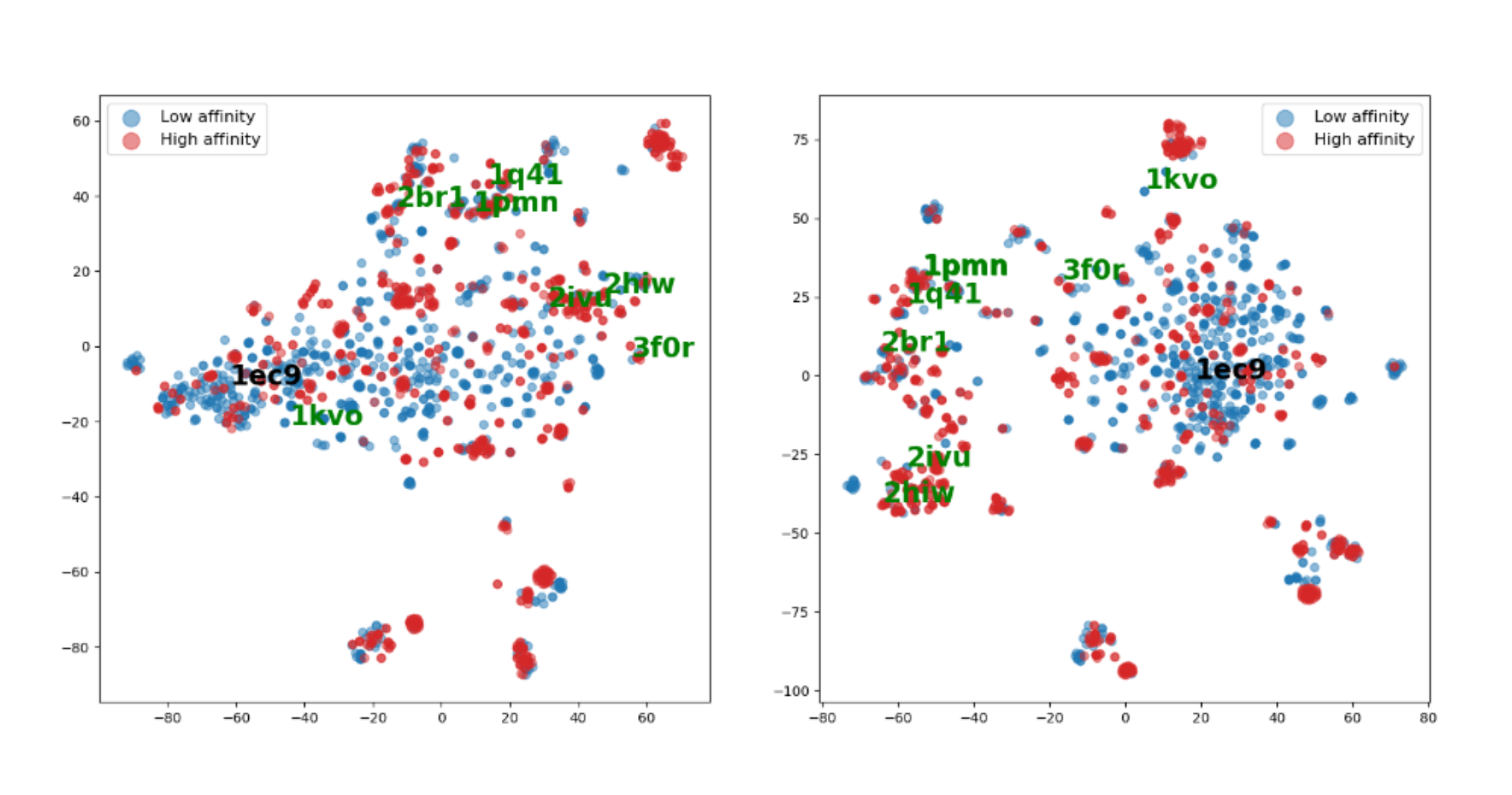} \caption{The T-sne plot of protein embedding of (a) ESM (b) ESM-PPI. Proteins are annotated with druggability which is black text for non-druggable and green text for druggable protein.
\label{fig:tsneprotein} }
\end{figure*}

Fig. \ref{fig:tsneprotein} shows the t-SNE plot of protein embedding with ESM encoder and ESM-PPI encoder using PDBBind cold-target test set. We also annotate the plot with druggability obtained from ‘NonRedundant dataset of Druggable and Less Druggable binding sites’ (NRDLD) dataset \cite{Krasowski2011DrugPred:Set}. In the PDBBind cold-target setting test set, the Glucarate Dehydratase (PDB:1ec9) is labeled as undruggable \cite{Krasowski2011DrugPred:Set}. We can observe the clear distribution of druggable and undruggable protein in the embedding space of ESM-PPI protein encoder. We hypothesize that the knowledge learned from PPI task can assist the druggability prediction. To verify this hypothesis, we use the ESM or ESM-PPI as the input for simple SVM model for druggability classification. We use the NRDLD dataset \cite{Krasowski2011DrugPred:Set} for training and validation. The result (Table \ref{tab:druggabilityNRDLD}) indicates that the knowledge learned from PPI task can help the model learn the druggability of protein, thus assisting the DTA task. 

\begin{table}[h!]
\caption{The result of druggability classification on NRDLD dataset \cite{Krasowski2011DrugPred:Set} using ESM and ESM-PPI features with a simple SVM model. The result shows that ESM-PPI clusters the druggability, thus improving SVM model performance. \label{tab:druggabilityNRDLD}}
\begin{center}
\begin{tabular}{c c c c c} 
\hline
Protein encoder & Precision & Recall & F1 & Accuracy \\ 
\hline
ESM&0.6803&0.8028&0.7349&0.6434\\
ESM-PPI&\textbf{0.6979}&\textbf{0.8742}&\textbf{0.7733}&\textbf{0.6869}\\
\hline
\end{tabular}
\end{center}
\end{table}

Looking back to the complex of resistant strain of HIV-1 protease (v82a mutant) with Ritonavir in Sec. \ref{sec:intro}, we compare the performance of model using only ESM and model with PPI transfer learning and ESM (ESM-PPI). The results in Table \ref{tab:1el8} shows that model with PPI transfer learning has a lower error rate than the model without PPI transfer learning. This implies that knowledge of protein interface and PPI integrates well into the DTA model.  

\begin{table}[h!]
\caption{The prediction of ESM and ESM-PPI model for the resistant strain of HIV-1 protease (v82a mutant) with Ritonavir.\label{tab:1el8}}
\begin{center}
\begin{tabular}{c c c} 
\hline
Protein encoder & Predicted affinity & Error \\ 
\hline
ESM&7.2532&1.1532\\
ESM-PPI&6.9038&0.8038\\
\hline
\end{tabular}
\end{center}
\end{table}

\subsection{\textbf{Integrating different types of CCI improves the DTA prediction model performance}}

The Chemical-chemical interaction in STITCH dataset \cite{Kuhn2008STITCH:Proteins} consists of not only interaction from experimental data but also interaction in a sense of similarity between activities or structure and literature text co-occurrence. The number of experimental data is only a small proportion of full CCI data. We hypothesize that not only the experimental interaction but also other types of interaction are useful for pre-training task. The results in Table \ref{tab:davisexponly} and \ref{tab:pdbbindexponly} show that pre-training with all types of CCI outperforms pretraining with only experimental data by a large margin. This suggests drug structure and activities similarity, as well as text co-occurrence can also provide useful information for DTA task.  

\begin{table}[h!]
\caption{The performance of the DTA model on Davis dataset with drug encoder pre-trained with only experimental interaction CCI and drug encoder pre-trained with all types of interaction available in the stitch STITCH dataset.\label{tab:davisexponly}}
\begin{center}
\resizebox{\columnwidth}{!}{%
\begin{tabular}{p{1cm} p{1cm} c c c c c} 
\hline
Protein encoder & Drug encoder & Pretrain & RMSE & Pearson & Spearman & CI \\ 
\hline
ESM&GIN-CCI&Full&\textbf{0.8755}&\textbf{0.575}&\textbf{0.5034}&\textbf{0.743}\\
&&Exp&0.98&0.3588&0.4275&0.707\\
&Chemberta-CCI&Full&\textbf{0.9146}&\textbf{0.5259}&\textbf{0.4485}&\textbf{0.7171}\\
&&Exp&1.07&0.346&0.3664&0.6769\\
ESM-PPI&GIN-CCI&Full&\textbf{0.8841}&\textbf{0.5564}&\textbf{0.4741}&\textbf{0.7299}\\
&&Exp&1.0398&0.3595&0.3706&0.6753\\
&Chemberta-CCI&Full&\textbf{0.9171}&\textbf{0.4906}&\textbf{0.4216}&\textbf{0.7034}\\
&&Exp&0.9181&0.4774&0.4087&0.6956\\
\hline
\end{tabular}
}
\end{center}
\end{table}

\begin{table}[h!]
\caption{The performance of the DTA model on PDBBind dataset with drug encoder pre-trained with only experimental interaction CCI and drug encoder pre-trained with all types of interaction available in the stitch STITCH dataset.\label{tab:pdbbindexponly}}
\begin{center}
\resizebox{\columnwidth}{!}{%
\begin{tabular}{p{1cm} p{1cm} c c c c c} 
\hline
Protein encoder & Drug encoder & Pretrain & RMSE & Pearson & Spearman & CI \\ 
\hline
ESM&GIN-CCI&Full&\textbf{1.3484}&\textbf{0.7236}&\textbf{0.7025}&\textbf{0.7603}\\
&&Exp&1.4053&0.6927&0.6638&0.7441\\
&Chemberta-CCI&Full&\textbf{1.3653}&\textbf{0.7059}&\textbf{0.6798}&\textbf{0.7498}\\
&&Exp&1.3816&0.7012&0.6696&0.7454\\
ESM-PPI&GIN-CCI&Full&\textbf{1.3379}&\textbf{0.7282}&\textbf{0.7039}&\textbf{0.7618}\\
&&Exp&1.4789&0.6672&0.6482&0.7351\\
&Chemberta-CCI&Full&1.3735&0.7009&0.6800&0.75\\
&&Exp&\textbf{1.3627}&\textbf{0.7112}&\textbf{0.6835}&\textbf{0.751}\\
\hline
\end{tabular}
}
\end{center}
\end{table}

\section{Conclusions and Future works}

In conclusion, migrating the cold-start problem in drug-target affinity prediction requires external knowledge from labeled and unlabeled data. Unsupervised learning such as language modeling learns the intra-molecule interaction and internal structure representation of the proteins and drugs from unlabeled data. The drugs and proteins representation are then imbued with inter-molecule interaction learned from similar tasks such as protein-protein interaction and chemical-chemical interaction. The protein-protein interaction can provides knowledge regarding protein surface, activity, druggability. The chemical-chemical interaction provides common pharmacological action, similarity in structure and targets. Combining both intra-molecule interaction and inter-molecule interaction information allows more robust drug and protein representation to deal with cold-start problem. In addition, interactions curated from different resources such as text mining are also useful for learning interaction knowledge.       

Protein-protein interaction is a complex interaction. Modeling the exact interaction between two proteins requires surface and structure information reflected in the protein encoding architecture such as graph or cloud points. Learning PPI with more dedicated architecture could potentially benefit not only DTA task but other tasks such as druggability as well.
\bibliographystyle{natbib.bst}
\bibliography{references.bib}
\end{document}